# Counting on Beauty: The role of aesthetic, ethical, and physical universal principles for interstellar communication.

**Guillermo A. Lemarchand**

## 1. Introduction:

The land lies sleeping under the enveloping mantle of night. Bright stars gleam like jewels from the velvet darkness. Beyond these points of celestial beauty, in depths frightening in their sheer immensity, lie realms of stellar glory. And beyond again—for there always must be a beyond—the Milky Way trails its tenuous gown of stardust across the heavens. Billions of planets, stars, and galaxies, dancing in a cosmic symphony.

Vast and old beyond understanding, the universe makes us feel lonely. From our earliest days, humans have strongly sensed that this endless majesty is too immense to be contemplated only by terrestrial intelligence. One thread that links the ancient Greek philosophers with modern space scientists is the desire to know whether there are other inhabited worlds with sentient creatures with whom we are share the vast beauty of the cosmos. It forces us to ponder the ultimate significance, if any, of our tiny but exquisite life-bearing planet, and to long for the knowledge that somewhere out there, someone like us is looking upward toward the heavens and having similar thoughts.

The *Principle of Mediocrity* assumes that anything seemingly unique and peculiar to us is actually one out of many and is probably average. By this view, the Earth and humans do not occupy a privileged position in the universe. The history of evolution of matter and life on Earth should be typical under the same universal physical laws and similar environmental conditions. Even not expecting the same evolutionary paths, we assume that the sequence of events may have similar patterns: the origin of life, the appearance of complex life; the emergence of intelligence, the development of culture and technology, and membership in the community of galactic intelligence.

For the first time in human history, we are testing the hypothesis that advanced civilizations exist elsewhere. Our search is based in the assumption that any advanced intelligent species must develop certain technological tools to explore their cosmic environment. The use of advanced technologies for that exploration would generate signals so different from any natural cosmic source, that if we detect them, we will be able to identify them as artificial. Behind an artificial interstellar signal, there must be some intelligent creature with the capacity to solve very complex problems.

How should we build a common language to exchange information and wisdom across the galaxy? Would such a hypothetical universal language, or *Lingua Cosmica*, let a civilization communicate its accumulated knowledge, as well as the differing perceptions, sensations and emotions of its species? We will explore these questions in the following essay.

## 2. Epistemic bases for an interstellar language

We introduce the term "cognitive map" to clarify the concept of paradigms, often discussed by philosophers of science (e.g., Kuhn, 1962). Cognitive mapping refers to the process by which an organism makes representations of its environment in its own processing system.

Features of the physical environment are only some of the significant parts of human cognitive maps. Our cognitive maps must represent such varied objects as living beings and their behavior, linguistic abstractions, aesthetic judgments, and ethical values.

With the emergence of symbolic language, environments can be described in words. Not limited to mapping only sensory perceptions, language permits the mapping of concepts as well. With languages and their rules, one can reconstruct events that occur in time. Through language, our species learned how to cooperate in space and time. To create such an interstellar language, we must propose a set of conjectures to transform our human cognitive maps into cognitive universals.

A *Lingua Cosmica* might provide an instrument for interspecies cooperation. From an ethical point of view, the last may be possible as a consequence of some sort of *Principle of Universal Fraternity*, in which advanced intelligent societies choose to help other species in order to maximize the development of potentialities for all the expressions of life in the universe.

Benjamin Whorf (1897-1941) showed that particular languages, through their grammatical structures, influence the kinds of thoughts people have, their ways of perceiving the world, and their ways of conceiving themselves. The process of learning a language recreates in a new generation the habits of mind, categories of interpretation, and patterns of behavior that typified their society in the past. The acquisition of language is vital to forming human cognitive maps, for it opens up a new dimension in cognitive structures. Every human language maps the environment of its native speakers, with linguistic information accumulating over many generations. Each language, therefore, reflects the particularities of the environment in which its speakers live. The Eskimos, for example, have more than twenty words for snow, while people living in the tropics can get alone with one or none. An extension to different galactic environments may present a strong incommensurability problem for exchanging information based on different cognitive maps.

Alien intelligence will be embodied in different forms, and it will have different needs, senses, and behaviors. Extraterrestrials could inhabit environments in which neither science nor technology is needed for survival. The science of an alien civilization would reflect their idiosyncratic ways of they perceiving nature, as funneled through their particular evolutionary history in environments quite different from those encountered by the precursors of humans. If we, and they, use widely differing cognitive maps, we may never be able to distinguish any of their manifestations of intelligence.

SETI researchers believe that the basic principles of our science and the science of extraterrestrial beings should be fundamentally the same, and we should be able to communicate with them by referring to those things we share, such as the principles of mathematics, physics, and chemistry. This view assumes that there is only one way to conceptualize the laws of nature. Consequently, mathematics and the language of nature should be universal.

In contrast, Kuhn (1962) envisioned situations in which two theoretical conceptions are so different that disputes between them cannot be resolved. In such situations, the two traditions are said to be *incommensurable*. The problem, according to Kuhn, is that each conception of science uses its own method to evaluate its theories. Moreover, theories themselves affect the very observations that are used to support scientific paradigms. Kuhn supported his analysis with case studies from the history of scientific revolutions. For example, he examined the transition from Ptolemaic to Copernican astronomy, as well as the shift from Newtonian mechanics to special relativity.

Similarly, the standards of evidence, interpretation, and understanding dictated by the terrestrial cognitive map on the one hand, and by an extraterrestrial cognitive map on the other hand, may be so different that mutual understanding may be impossible. To comprehend a radically different scientific view, one must adopt a wholly new conception of the world, complete with its own distinctive standards of rationality. With no significant theory-independent standards of rationality, there would be no basis for communication.

Two different types of incommensurability may limit the exchange of intelligible information between human and extraterrestrial cultures. In methodological incommensurability, no rational method for exchanging intelligible information can be found because of differences in observation and scientific method. In semantic incommensurability, the use of the same term in two different cognitive maps or paradigms has clear different meanings according to the different epistemological worldviews. For example, instances of the term “mass” in Newtonian and relativistic mechanics are incommensurable.

But perhaps such extreme epistemological skepticism may be unwarranted in SETI. We need not anticipate all possible aliens, but only those similar enough to ourselves to create radio or laser transmitters. The only way we have to find evidence of intelligent extraterrestrials is through the detection of artificial signals generated by technological activities, and many sorts of cognitive universals can be surmised by considering the pragmatic requirements for building equipment to send and receive electromagnetic signals. Technological aliens will have evolved thought processes and communication strategies that will match our own well enough to allow us to comprehend them.

Since the early works of Johann Becher (1635-1682) and Gottfried Leibnitz (1646-1716), it has generally been agreed that mathematics is based on cognitive universals that cut across cultural and linguistic boundaries between humans. It is also where we are likely to find cognitive universals that can be used to communicate with extraterrestrials. The argentine writer Jorge Luis Borges (1952) wisely noted that the main limitation of any universal language design is the artificiality of the classes and categories in which the knowledge is organized. As an example he mentioned certain Chinese encyclopedia entitled ‘Celestial Empire of Benevolent Knowledge’: “In its remote pages it is it is written that the animals are divided into: (a) belonging to the emperor, (b) embalmed, (c) tame, (d) sucking pigs, (e) sirens, (f) fabulous, (g) stray dogs, (h) included in the present classification, (i) frenzied, (j) innumerable, (k) drawn with a very fine camelhair brush, (l) etcetera, (m) having just broken the water pitcher, (n) that from a long way off look like flies.”

Minsky (1985) considered that all intelligent problem-solvers will be subject to the same ultimate constraints: limitations on space, time, and resources. For intelligent life forms to evolve powerful ways to deal with such constraints, they must be able to represent the situations they face, and they must have processes for manipulating those representations. That is, they must have complex cognitive maps and languages. Every intelligence must develop symbol-systems for representing objects, causes, and goals, and for formulating and remembering the procedures it develops for achieving those goals. During its evolution, each species will eventually encounter certain very special ideas that are much simpler than other ideas with similar uses. Thus, we can expect extraterrestrials to know about arithmetic, causal reasoning, and optimization processes.

## 3. Searching for a *Lingua Cosmica*

The origin of language is clearly related to cognitive capability. Chomsky claims that language has a biological basis, the constraints of which lead to certain universals in human languages. Can we extend our terrestrial cognitive map to create a symbolic system to exchange information with alien cultures? Can we assume that a Principle of Mediocrity holds, in which the patterns found in the emergence and evolution of human language should be average in the galaxy? If there is an evolutionary convergence among these paths, an analysis of the origins and development of human language may shed some light on the emergence of extraterrestrial languages, which in turn might help us create a more intelligible *Lingua Cosmica*.

In significant ways, our speculations on the likely forms of interstellar messages may be limited by our own experiences as linguistic beings—-whether or not these experiences are truly universal. Consider, for example, how lessons learned from child development on Earth might affect our expectations about any messages we might receive from extraterrestrial intelligence.

Piaget (1955) observed that children spend much time talking to themselves. Even conversation *between* children is often a “collective monologue.” Although two or more children may appear to be responding to one another, in reality they may only be pursuing their own monologues, without taking into account the reactions of others (Merleau-Ponty, 1973). As Piaget noted, early speech is egocentric. Dialogue appears to be secondary. Might this terrestrial fact influence the sorts of messages we anticipate from extraterrestrial civilizations?

For decades, scientists have been speculating about the characteristics of alien contact. Since the early works in the late 1950s, the predominant view has been that an advanced technological extraterrestrial society would send us some kind of *Encyclopedia Galactica*. In it, they would tell us everything about other worlds, the laws of nature, and eventually, recipes to avoid self-destruction through the misuse of our technology and the misconduct our species. During all these years there has been a basic assumption that signal from ETI would be a one-way message. Extraterrestrials will speak, and we will listen; they will teach, and we will learn. The vast distances of interstellar space, with information being sent via electromagnetic waves, implies timescales of hundreds or thousands of years for each exchange. Only those societies with long life expectancies would be patient enough to start a dialogue, and not settle for only monologues.

There is a close analogy between expectations about interstellar communication and standard theories about the emergence of language in humans. Both start with a monologue. Just as early human speech is egocentric, our early attempts to design interstellar languages have been homocentric.

## 4. Underlying Dimensions of a *Lingua Cosmica*

The acronym SETI is in some sense misnomer. It is impossible to detect any evidence of intelligence over interstellar distances, but we may find traces of artificial sources generated by technological activities. So, the research program is focused in the Search for Extra Terrestrial Technological Activities or SETTA (Lemarchand, 1992).

In SETTA, as well as in conventional astronomy, the cosmic distances are so huge that the researcher can only observe from afar, entirely dependent on carriers of information that traverse the universe. According to our cognitive map of physical laws, the information carriers are not infinite in variety. All information we currently have about the universe beyond the solar system has been transmitted to us by means of electromagnetic waves (radio, infrared, optical, ultraviolet, x-rays, and $\gamma$-rays), cosmic ray particles (electrons and atomic nuclei), and more recently, by neutrinos (Lemarchand, 1992, 1994a).

Table 1 summarizes, according to our cognitive map of the universe, the possible information carriers that may let us find the first evidence of an extraterrestrial technological civilization. Applying Imre Lakatos epistemological approach for the SETTA research program, in the *hard core* of this particular interpretation of reality are two hypotheses: 1) the Principle of Mediocrity and 2) that the laws of nature that characterize the information carriers are universal. That is, the laws of physics are independent of any particular position in space or time. All modern astrophysics assumes the last hypothesis and considers it irrefutable.

We continue testing other physical theories that constitute a *protective belt* around the hard core. In our case, the protective belt includes the complete list of potential information carriers. We then see which aspects of the protective belt we keep, and which we need to discard to protect our core assumptions about the universality of the laws of nature and the Principle of Mediocrity. Regardless of what information carrier is used, in order to detect that a signal exists and that is of artificial origin, we define a *positive heuristic*, for the selection of the appropriate information carrier following these requirements: the received signal must significantly exceed the natural background count, the signal must exhibit some property not found in natural sources, required the least radiated power, not be absorbed by the interstellar media or planetary atmospheres, not be deflected by galactic fields, permit efficient generation and detection, travel at high speed. According to our cognitive map the electromagnetic waves satisfied all the requirements. Using a *negative heuristic*, we may reject as useful for interstellar communication several of the information carriers, such as graviton astronomy, pneumatic astronomy, and particle astronomy, with the possible exception of neutrinos (Lemarchand, 1992, 1994a). A similar approach, applying both heuristics, was recently used when the SETI Institute identified its research priorities for the next two decades (Ekers et al., 2002).

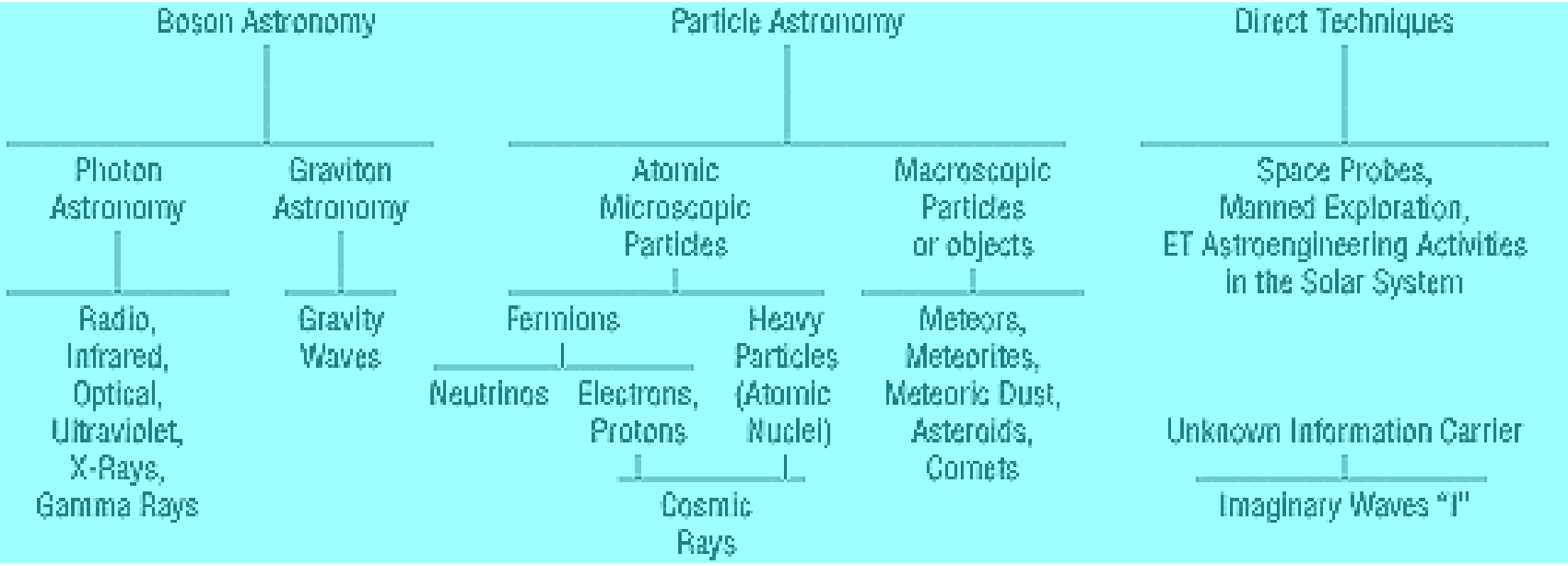

**Table 1: The humankind's cognitive map –or worldview- about the universe outside the Earth was obtained using the following list of physical information carriers. Applying a positive and negative heuristic, we may see which of these *information carriers* are the best ones to obtain evidences of extraterrestrial technological activities beyond the Earth. This table includes a separate item called *"Imaginary Waves"* to call the attention to the fact that there should be several unknowns physical laws that may be more efficient that the present ones for interstellar communication.**

Each observational technique acts as an information filter. Only a fraction (usually small) of the complete information can be gathered. The diversity of these filters is considerable. They strongly depend on the technology available at the time. Consequently, cognitive maps about the hypothesized characteristics of an extraterrestrial message changed over time. At the beginning of the twentieth century, scientists, including Einstein, speculated about using light-signals to contact Martians. Later on, Marconi, Tesla, and Todd proposed the use of radio waves to do the same. In the late 1950s, Cocconi, Morrison, Drake, and Shklovskii started the modern radio astronomical search for interstellar communication, while in the last ten years we have initiated the nanosecond search for interstellar laser pulses. These few examples show how the assumptions about the best means of interstellar communication have shifted as terrestrial technology has evolved.

The best search strategy would be based on a set of cognitive universals known by any extraterrestrial culture. These cognitive universals would reflect the essence of the physical structure of a hypothetical extraterrestrial artificial signal; they would also provide the fundamental principles for constructing a *Lingua Cosmica*.

Obviously, no matter how careful we are in attempting to identify universal concepts, we will have biases due to our culture (*ethnocentrism*) and our human nature (*homocentrism*). In fact, it may be impossible to avoid these biases, and we may never know how they affect our ability—or even make it impossible—to recognize, decode, and respond to communications from extraterrestrial intelligence.

Our human cognitive map of the universe and our best judgments about appropriate interstellar communication technology can be described by a cosmic haystack composed by several dimensions. The large number of possible combinations of the different dimensions in the cosmic haystack requires the development of a methodology to find cognitive universals or the key for a *Lingua Cosmica*.

We use positive and negative heuristics to identify optimal scientific, technological, aesthetic, and ethical principles for interstellar communication. Applying both heuristics to the information carriers presented in Table 1, we assume that only electromagnetic waves satisfy the criteria needed to exchange information across interstellar distances.

The nature of interstellar communication can be described through the complex interplay of the different dimensions of the cosmic haystack. We can locate extraterrestrial signals in time and space by noting when the signal was sent, by identifying the distance to the signal's source, and by locating the source in the sky map in terms of two coordinates that astronomers call right ascension and declination. Just as the source of the signal can be located in space and time, so too can the nature of the signal itself. Assuming that electromagnetic waves are the best information carriers, according to our cognitive map, we may explore the physical properties of this type of signals. Especially relevant characteristics of the signal include its frequency, width, polarization, the method by which it is modulated to bear information, and the rate at which information is transmitted. The content of any information borne by the signal is influenced by the kind of information, the meaning to be conveyed, and the encoding scheme. The interplay between these twelve dimensions suggests many possible approaches to interstellar communication, each based on cognitive constraints plausibly recognized by both humans and extraterrestrials.

This approach builds upon Dixon's (1973) anti-cryptography principle, which suggests that an interstellar signal meant to initiate contact must be designed and transmitted to maximize the possibility of discovery, both by intentional searches or serendipitous discovery, for example, by astronomers studying anomalous observations. To facilitate the discovery of the signal, the sender must minimize the number of unknowns in the dimensions listed above. This principle helps generate positive and negative heuristics, which in turn help overcome differences between terrestrial and extraterrestrial cognitive maps.

To identify the characteristics of optimal signals, we can start by building a matrix that relates all the possible alternatives of the twelve dimensions of the cosmic haystack, against the restrictions imposed by scientific, technological, aesthetic and ethical constraints. This will result in a multiplicity of proposals, each satisfying the anti-cryptography principle. These scientific and technological proposals draw upon plausibly universal cognitive constraints to identify the best heuristics for designing interstellar messages that may avoid many of our homocentric biases.

Table 2 relates the twelve dimensions of the cosmic haystack with a set of scientific, technological, aesthetic, and ethical constraints, for developing search strategies based on cognitive universals that provide a foundation for constructing anti-cryptographic messages. Far from being complete, Table 2 represents a first attempt to design a holistic methodology to find cognitive universals. Due to the lack of space in this essay, this table shows only the last three dimensions of the cosmic haystack (code, semantics, and kind of information content). The complete version of the table with a full analysis of all of twelve dimensions and thirty-four proposals that follow from the matrix may be obtained from the author.

## 5. Ethical constraints:

In order to improve the lifetime of a technological civilization it is impossible to have superior science and technology and inferior morals. This combination is dynamically unstable and we can guarantee a self-destruction within the lifetimes of advanced societies ($10^5$ to $10^6$ years?). The society of Earth has often been described as being in a stage of technological adolescence. Still struggling with environmental and security issues that threaten the continued existence of life on our planet, we might wonder whether advanced extraterrestrials would have similar evolutionary paths. Would extraterrestrials show the same disparity between technological development and social maturity that we see on Earth?

When any civilization reaches its technological adolescence, it must either be or become socially responsible, or face self-annihilation. Otherwise, global extinction would be very likely, with a life span too short to engage in a long-term interstellar dialogue.

If individuals within an extraterrestrial society can learn to co-exist peacefully with one another, might they not extend their attitude of benevolence to civilizations on other planets as well? Their own success at surviving the threats of planetary extinction would teach them vital lessons about respect for others that would likely be extended beyond their own world.

Cognizant of the uniqueness of each planet's evolutionary history, we might expect advanced extraterrestrials to avoid interfering with the development of less advanced species, resulting in a galactic quarantine guided by *Lex Galactica* (Lemarchand, 2000). A consequence of this galactic quarantine may be the so-called *Principle of Universal Fraternity*, which may be used by extraterrestrials to design the kind of information content of an interstellar message and to define a policy to help other species to maximize the development of its own potentialities.

If such constraints on interspecies relations hold, what implications would it have for SETTA? First, we might expect that young civilizations would be given only very limited information about science and technology, for fear of providing tools for self-destruction.

For civilizations more advanced than ours, such constraints might not be as rigid. Older civilizations, having withstood the test of time, might be able to assimilate information from other worlds without the same threat to their stability and longevity.

Earthlings would be hard-pressed to provide evidence that we are ready to learn about advanced technologies. Our unintentional signals, leaking from the Earth's radio and television stations, announce our existence to other worlds only to relatively nearby stars. Beyond about 70 light-years from Earth, intelligent beings will still not have received signs of terrestrial technology. As a result, they would not be able to send us age-appropriate interstellar messages. Not wanting to share potentially dangerous information with us, we might expect signals from other worlds be relatively impoverished in information content.

## 6. Symmetry as an aesthetic cognitive universal for art and science

Lemarchand and Lomberg (1996) introduced the use of aesthetic principles to evaluate search strategies for interstellar contact. We called attention to some properties of symmetry that may be used, not only in the coding and meaning of a message, as well as other aspects of signaling. We also analyzed the universality of particular patterns of aesthetics, such as the golden section and the radial symmetry in nature, art, and science.

Symmetry often indicates the deep structure of things, whether they are natural phenomena or creations of artists and scientists. We find symmetry pleasing, and its discovery is often surprising and illuminating as well. As it was previously stated this in one of the hypothesis behind the *hard core* of the SETTA research program. Indeed, the assumption that laws of nature are universal is based on notions of spatio-temporal symmetries. For example, Earman (1978) identified several epistemological characteristics of universal laws of nature: they do not make reference to specific spatio-temporal locations or coordinates, they are unrestricted in space and time, and they are invariant under spatio-temporal translations. Thus, symmetry is a necessary condition to consider a law of physics universal.

Symmetry also serves as a bridge between art and science. Several prominent scientists consider issues of aesthetics and symmetry when choosing between theories. Inspire by the views of Albert Einstein (1879-1955) and Hermann Weyl (1885-1955), more than any other modern physicist Paul Dirac (1902-1984) was preoccupied with the concept of mathematical beauty as an intrinsic feature of nature and as a methodological guide for its scientific investigation. Speaking about the aesthetics of science, he noted that “A physical law must possess mathematical beauty” and “a theory with mathematical beauty is more likely correct than an ugly one that fits some experimental data.” These views about the aesthetics of science and truth may provide an additional constraint for designing interstellar messages.

Symmetries also represent a kind of invariance under transformation, which implies that symmetrical things contain redundancies. Thus, sometimes information originates in the breaking of symmetries. Information as well as beauty, it would seem, are products of the interplay between symmetry (thesis), asymmetry (antithesis), and dissymmetry (synthesis).

The fact that the concept of symmetry has roots in both science and art make it useful to build bridges between various disciplines. Ancient Greeks realized that symmetry represents some of the objective and measurable aspects of beauty, and they represented symmetry by proportion and asymmetry by the lack of proportion.

Eventually, proportion became a general expression referring to the ratio of two magnitudes (a/b) or a set of ratios (a/b, c/d, ...). In each case, the pairs of magnitudes in a proportion are associated with the parts and the whole of an object. The concept of proportion is important in aesthetics because it can be quantified and used for comparison. However, the original meaning of proportion was not just a ratio or a set of ratios, but the equality of two or more ratios.

The golden section, discussed under a variety of names, has been well-known in geometry since ancient times. Later it also gained some importance in art as a proportion that symbolizes the harmony between the parts and the whole, and which is associated with organic shapes, from the DNA molecule to the human body. The term golden section refers to the division of a line segment or time interval into two parts, so that the ratio of the smaller part to the larger one is the same as the ratio of the larger one to the whole. The golden section, then, represents the harmonic resonance between the parts and the whole. Kepler (1571-1630) provides an excellent example of how highly the golden section has been regarded: "Geometry has two great treasures: one is the Theorem of Pythagoras: the other the division of a line into extreme and mean ratio. The first we may compare to a measure of gold; the second we may name a precious jewel."

The Greeks had already noticed that at least three terms are needed to express a proportion; such is the case of the continuous proportion $\frac{a}{b}=\frac{b}{c}$. But the problem can be simplified by using only two of the original terms, and expressing the third term as the sum of the first two terms. For example, *(a/b) = (b/c)* can be simplified by making *c = (a + b),* so the continuous proportion becomes $\frac{a}{b}=\frac{b}{(a+b)}$. Reshuffling the terms in the last equation gives us $b=a\frac{(1+\sqrt{5})}{2}$. The constant of proportionality (the numerical part of this equation) is called the golden section, represented by the Greek letter $\phi$, which is approximately equal to 0.61803..., or expressed as a fraction, $\frac{1}{1.61803....}$. The use of this proportion as the foundation for aesthetically pleasing images can be seen throughout this volume, in the drawings by artist Steve Deihl that introduce each section of the book.

The golden section has a great variety of interesting mathematical properties, some of which are not immediately obvious. For example, the Italian mathematician Leonardo Pisano (1170-1250), known as Fibonacci, described a sequence of numbers that starts with two 1's, with each subsequent term being the sum of the previous two terms: 1, 1, 2, 3, 5, 8, 13, 21, 34, 55, 89, 144, 233.... Fibonacci presented this series in the context of a very artificial exercise about the growth of the number of rabbits from one generation to the next, without providing further details. Later it was discovered that the ratios of any two neighboring Fibonacci numbers approximate the golden section number, with the approximation becoming increasingly accurate the farther into the series one goes. For example, the ratio between the two terms 8 and 13 is 0.61538.... An even more precise approximation of ϕ can be found by looking at the ratio of two successive terms later in the series. For example, the ratio of 34 and 55 is 0.61818....

The golden section makes many appearances in geometry. For example, imagine we start with a Pythagorean pentagram having all sides of equal length and all inside angles identical. We can then draw a five-pointed star inside this pentagram, with each line of the star bisecting two other lines that make up the star. These bisected lines are cut into pieces that have lengths related by ϕ. That is, one segment is 0.61803... times as long as the other segment. In this case, we also see an example of self-similarity, also seen in the more general class of mathematical objects called fractals. Just as the length of the shorter segment is 0.61803... times the length of the longer segment, so too is the longer segment 0.61803... times the length of the entire line before being cut. The golden section is also found in right triangles with sides having the ratios of 3:4:5 and in many other areas of mathematics.

### 6.1. The golden section in the physical world

The golden section and the Fibonacci series appear in the real world in many physical and biological phenomena. Examples from the physical world include:

- If a beam of light falls upon two sheets of glass that are in contact, there will be multiple reflections. The number of rays that will emerge from the sheets of glass depends on the number of times the ray is reflected. The number of emergent rays is always a Fibonacci number.
- A spiral can be formed along the ends of adjacent, perpendicular lines, if the proportion of the lines to one another is ϕ. This lovely spiral is seen in some of the most handsome spiral galaxies, such as M51, M101, and M74.

### 6.2. The golden section and life patterns

The golden section may also be seen in biology in population growth of successive generations, in the genealogy of the drone bee, and in the average value of leaf divergence measured on the stems of plants.

The number of petals on many common flowers is a Fibonacci number. For example, the iris has 3 petals, the primrose has 5, the ragwort has 13, and the daisy has 34. Plants with composite floret patterns, like the sunflower, the pinecone, and the pineapple show overlapping clockwise and counter-clockwise spiral patterns. In the sunflower, there are 21 spirals in clockwise direction, and 34 in the counter-clockwise direction; the pinecone has five spirals 5 clockwise and 8 counter-clockwise; in the pineapple, the spirals number 8 clockwise and 13 counter-clockwise.

The same pattern that describes the shape of spiral galaxies, based on the golden section, also appears in the curve of animal horns and in the seashell of *Nautilus pompilius*, the chambered nautilus.

### 6.3. The golden section in human arts

Beautiful patterns based on the golden section have been created by human artists for 2500 years. The Greek sculptor Phidias (5th century BCE) used it, and the proportion of the height to the length of the facade of the Parthenon being a close approximation to $\phi$. In the two millennia since the Parthenon was designed, $\phi$ has continued to be a touchstone for generations of graphic and design artists and architects from Leonardo (1452-1519) to Mondrian (1872-1944).

We cannot say why this "golden" proportion occurs in processes as seemingly unrelated as galactic shapes and growth patterns of flowers, horns, and shells. But for whatever reason, it seems to be deeply embedded in the mathematical description of phenomena throughout the real universe. It is reasonable to assume that other intelligences studying the cosmos will have discovered this proportion and patterns based upon it as well. The golden section may be a cognitive universal that has a particular capacity to connect parts with a whole.

Behind this simple idea there is a very profound concept. Our traditional cognitive maps divide the whole into parts in order to study them and to make the universe intelligible for us. A different epistemological approach may be the study of the wholeness and its implicate order (Bohm, 1980). Some decades ago, a group of physicists started using these ideas to explore the quantum world. Unfortunately, our language is inadequate to describe the properties of the whole, and so too are our cognitive maps inadequate. Perhaps some advanced civilizations have reached this point of development. The golden mean is a simplest expression of these ideas.

## 7. The application of the golden section as a cognitive universal for interstellar communication codes, semantics, and interstellar artistic works

Table 2 shows various criteria for choosing cognitive universals to compose interstellar messages. As an example, here I propose a way to convey aesthetic meaning through a three-dimensional message, allowing us to transmit a hologram or a short two-dimensional movie.

The golden section is a mathematically incommensurable proportion, what means that the exact number $\phi$ has an infinite number of decimals. One never can use an exact golden section of time-intervals in a temporal sequence of information bits sent by an electromagnetic interstellar message that is organized on equal units of intermittent signals. Here all marked time intervals (length of pulses, duration of special instructions, etc.) can be measured by integer times of the corresponding unit, therefore all the possible ratios are rational. Anyway, it is possible to express the approximate value of the golden section in an interstellar message making a dimensional transformation from space to time, and applying the Fibonacci sequence.

The composer Iannis Xenakis (1971) explored this idea, using magnetic recordings of music. Similarly, the Russian director Sergey Eisenstein (1898-1948) used the golden section when composing the layout of important scenes in his movie *Bronenosets Potemkin* (*The Battleship Potemkin,* 1925). Such precise measurements, which are secrets of the composer or movie director, are very rare and can hardly be detected by the audience.

The most parsimonious proposal for using these cognitive universals in an interstellar message may be the following: the calling signal may start with the first 13 numbers of the Fibonacci series (1, 1, 2, 3, 5, 8, 13, 21, 34, 55, 89, 144, 233) encoded in a binary form, with the sequence repeated over and over again. This series will provide a clear criterion of artificiality and also a pattern related to aesthetics and life.

We propose transmitting three-dimensional holograms, using an aesthetic cognitive universal. To do so, we describe the construction of a golden cuboid, in which there are sides of three different lengths, and these lengths are related to one another through the golden section. Using some elementary algebra, it is easy to show that the dimensions of a cuboid (rectangular parallelopiped) of unit volume can have edges with lengths (*a, b, c*) related to the golden section, if we let the shortest side (*a*) be $\phi$ = (0.61803...) units long, the next longest side (*b*) be 1 unit long, and the longest side (*c*) be the reciprocal of $\phi$ (or $\phi^{-1}$ = 1.61803...) units long. When we calculate the volume of such a golden cuboid by multiplying the three sides ($\phi \times 1 \times \phi^{-1}$), we note that the volume is 1.

After the calling signal with the first 13 terms of the Fibonacci series, we may introduce a more subtle information message coded as a sequence of two level signals (on-off, 0-1, etc.) with 2,986,128 different time intervals. This is a representation of a three-dimensional golden cuboid message constructed as the product between three special Fibonacci numbers: 89 x 144 x 233 = 2,986,128. The last number will also place a limit to the equivalent number of bits for each golden cuboid message unit.

We have proposed these three numbers because of their special qualities. The number 144 is the only square number of the Fibonacci series; moreover, it lies between two prime numbers that are also contiguous members of that series (89 and 233). Due to the particular character of the number 144 in the series, this number may be taken as a second order basic unit of measurement. In this way we may proceed by dividing each of the three numbers in the sequence by 144, yielding normalized numbers with $a = 89/144 \approx \phi$, $b = 144/144 = 1$, and $c = 233/144 \approx \phi^{-1}$. The result is a three-dimensional aesthetic module, in which we can insert representations of solid objects or a sequence of 144 two-dimensional images to create a short movie. Each of these 144 images will be have the dimensions of the two prime numbers 89 and 233.

We may also encode more complex information reflecting some interesting aesthetical and mathematical properties of this *golden cuboid*, such as: (1) the ratios between the areas of the faces follows the relation $\phi:1:\phi^{-1}$, (2) the total surface area of the cuboid is $[3 \times (\phi + 1 + \phi^{-1})] = (6 \times \phi)$, (3) four of the six faces of the cuboid are also golden section rectangles, which may be use to order the temporal sequences of images, e.g., by creating a movie that unfolds in the direction perpendicular to the non-golden faces (4) The ratio of the area of the sphere circumscribing the cuboid to that of the cuboid is $2\pi:3\phi$, an interesting relation between primes and incommensurable numbers, and finally, (5) the cuboid also shows a self-similar or fractal pattern: if two cuboids of square cross section ($\phi^{-1} \times \phi^{-1}$) are cut from the golden cuboid, the edge lengths of the remaining cuboid are in the same ratio as those of the original cuboid, i.e., $1:\phi^{-1}:\phi^{-2}$ =

$\phi:1:\phi^{-1}$. This results in a new golden cuboid, $\phi^{-3}$ times the original size, providing a property may be useful to encode a higher level of information.

The use of the golden cuboid is only one of the many suggestions in this volume for designing interstellar messages. The primary purpose of this essay, however, was to call attention to some of the epistemological principles behind those proposals. In the process, we have also proposed a methodology for choosing cognitive universals restricted by scientific, technological, aesthetic, and ethical principles to provide a holistic framework for considering interstellar communication.

**References:**

1. Borges, J.L. (1952), El Idioma Analítico de John Wilkins, in *Otras Inquisiciones 1937-1952*, Buenos Aires: Sur.
2. Bohm, D. (1980). *Wholeness and the Implicate Order*. London: Routledge.
3. Dixon, R. (1973). A Search Strategy for Finding Radio Beacons, *Icarus, 20,* 187-199.
4. Kuhn,T. S.(1962). *The Structure of the Scientific Revolutions.* Chicago: University of Chicago Press.
5. Earman, J. (1978). The Universality of Laws, *Philosophy of Science*, *45* , 173-181.
6. Ekers, R.D, Cullers, D.K. Billingham, J. & Scheffer, L.K. (Eds.). (2002). *SETI 2020: A Road-map for the Search for Extraterrestrial Intelligence*. Mountain View: SETI Press.
7. Lemarchand, G.A. (1992). *El Llamado de las Estrellas*, Buenos Aires: Lugar Científico.
8. Lemarchand, G.A. (1994a). Detectability of Extraterrestrial Technological Activities, *SETI-Quest*, *1 (1),* 3-13.
9. Lemarchand, G.A. (1994b). Passive and Active SETI Strategies using the Synchronization of SN1987A, *Astrophysics and Space Research, 214*, 209-222.
10. Lemarchand, G.A. (2000), Speculations on the First Contact: Encyclopedia Galactica or the Music of the Spheres? In Allen Tough (ed.), *When SETI Succeeds: The Impact of High-Information Contact* (pp.153-163). Bellevue: The Foundation for the Future.
11. Lemarchand, G.A. and Lomberg, J. (1996 July) Is There any Universal Principles in Science and Aesthetics that could help us to Set the Unknown Parameters for Interstellar Communication? Paper presented at *International Astronomical Union Colloquium 160*, Capri, Italy.
12. Merleau-Ponty, M. (1973). *Consciousness and the Acquisition of Language*. Evanston: North-western University Press.
13. Minsky, M. (1985), Why intelligent aliens will be intelligible?, in E. Regis (ed.) *Extraterrestri-als: Where Are They?*, pp.117-128, Cambridge: Cambridge University Press.
14. Piaget, J. (1955). *The Language and Thought of the Child*. Cleveland: Meridian.
15. Xenakis, I. (1971), *Formalized Music*, Bloomington: Indiana University Press

# Table 2: SETTA Matrix*:*
## *Space of Configuration* for an interstellar communication channel

(This is the complete version of the Table in the book only an abridge version will be printed)

| Message's Variable | Universal Convergence Proposals | Constraints | | | |
|---|---|---|---|---|---|
| | | **Scientific** | **Technological** | **Aesthetical** | **Ethical** |
| **Spatial Coordinates:**<br><br>**A.Radial Distance:*(R)*, B.Right Ascension:*(α)* C.Declination:*(δ)*** | 1.Target Search (e.g., nearby stars, cosmic masers, etc.) | **A.**Using the *Principle of Mediocrity*, civilizations may evolve in suitable planets around solar type stars. If we assume that a distribution of $10^6$ hypothetical civilizations is homogeneously distributed in the galaxy, this implies that the nearby ETI would be some hundreds light years away. The target search criteria would be determined by this type of assumptions.<br>Other target candidates are cosmic masers, that have the possibility to naturally amplify artificial signals. This means that with a modest output transmitting power it is possible to be detected in the whole galaxy.<br>A third kind of targets are those galactic objects that may serve as cosmic synchronizers (e.g. supernova explosions).<br>**B and C**: Positions of solar type stars, cosmic masers and other target candidates (e.g. cosmic spots synchronizers). | With a "modest" -in cosmic terms- technology less than 100 years old, we are able to detect artificial microwave signals transmitted from the center of our galaxy by a civilization with a similar level of development than our terrestrial one.<br>*Project Phoenix* and *The Allen Telescope Array* are in the frontier of our technical capability.<br>There are several optical searches around solar-type stars looking for nanosecond pulses. | | Any advanced civilization within a sphere of ≈ 70 light years (ly) away may detect our first radio signals and begin studying our level of ethical development in order to calibrate what kind of "information" would not be dangerous for us. Applying the *Lex Galactica* (*LexGal*) and the *Principle of Universal Fraternity (PUF)*, eventually, they would be able to share with us the first chapters of their *Encyclopedia Galactica.* |
| | 2. Full-sky survey | **A**. The rationality of full-sky surveys is based in the fact that it is more easier to detect those beacons that are intrinsically strongest, but remote, that those which are nearer but weakest. For every 300 civilizations transmitting at certain radio power, there is but one civilization which transmits signals ten times more powerful. The last one would be much easier to detect.<br>**B.** $0^o \le \alpha \le 360^o$<br>**C.**$-90^o \le \delta \le 90^o$<br><br>The galactic plane is the most interesting place to search due to the high density of stars, so the probability of discovery a strong signal there is higher. | SERENDIP from Arecibo, Parkes, Italy and the META II project from Argentina are the most sensible full-sky surveys. There also are some full sky amateur observations coordinated by the SETI League.<br>Paul Horowitz and colleagues, have recently started a full-sky survey of optical laser pulses with nanoseconds resolution.<br>There are some projects to develop full-time omni directional antennas, to look in all sky directions, all the time. | If an advanced civilization *(R>> 70 ly)* is beaming us a high information message, by applying the *LexGal* and the *PUF*, the message would probably be manifestation of art. For doing so they would have to match their aesthetic principles using some "cognitive universals" related with symmetry, asymmetry and dissymmetry properties (e.g. the use of proportions, golden section and any of the limited number of musical scales that the laws of nature allows). | Advanced galactic civilizations beaming beacon signals, would not include a high technological information content message, because they have no possibility to calibrate the level of ethical development of the recipient societies.<br>High information content would put in danger the continuity of the recipient civilization due to the possible misuse of the new technological information. So, for the case of *R >> 70 ly*, only beacons or eventually "artistic messages" would be expected (e.g. music, visual arts or an unexpected new branch of arts). |
| **Timing *(t)*** | 3.Long duty cycle | Very advanced civilizations (Kardashev's Type II and III) would transmit full-time omnidirectionally in order to facilitate the detection by every emergent galactic civilization | A Kardashev's Type II ETI would use the equivalent transmitting power of a star for interstellar communication ($10^{26}$-$10^{27}$ W).<br>A Kardashev's Type III ETI would use the energy resources of a whole galaxy ($10^{37}$-$10^{38}$ W).<br>The full-sky surveys at Harvard and Buenos Aires have ruled out the existence of | Even this strategy makes a high symmetrical assumption of transmitting the same beacon signal simultaneously to all sky directions, the engineering and ethical variables would ruled out this option. | Omnidirectional transmission would demand extraordinary energy resources for a society, as well as a high radio polluted environment. This would not only be irrational from an engineering but also from an environmental ethics point of view. |
| | 4..Nova and Supernova Ellipsoids for temporal Convergence and others similar proposals | Nova and Supernova galactic explosions may serve as synchronizers to coordinate direction in the sky with the arrival of that signals to the terrestrial detectors (Lemarchand, 1994b). | Type II and III ETI in the galaxy employing electromagnetic waves at λ=21 cm.<br><br>With a modest transmitting power it is possible to send call signals to other civilizations in that direction. It is also possible to generate fake pulsar signals in order to improve the odds of detection by galactic astronomers [see Lemarchand (1994b) for an example on how to build a message for a nearby star]. | The *SETI Ellipsoid* is a high symmetrical and beautiful solution for the problem of synchronization between transmissions and receptions of interstellar communication among unknown partners. | The transmitting civilization would not necessary know the exact level of ethical development of the recipients. Applying the *LexGal* and the *PUF*, would only transmit a beacon signal, just to call the attention, then after receiving the first reply would decide what chapters of the *Encyclopedia Galactica* are not dangerous for the recipients. |
| **Contact Frequency *(f)* or equivalent Wavelength *(λ)*** | 7. Microwave "magic frequencies" (e.g. λ = 21 cm; 18 cm, 19.5 cm, 11.7 cm, 6 cm; 13.5 mm; 4.3 mm, 2mm, 1.7 mm, 1.5 mm, etc)<br><br>8. All microwave frequencies between 400 and 10,000 MHz<br><br>9. Flash-light, no specific frequency for interstellar laser pulses. | For example λ = 21 cm corresponds to the hydrogen emission line wavelength. Hydrogen is the most abundant and simple element in the universe. Any civilization that would like to study the distribution of galactic matter would need to develop radio astronomical studies of the galaxy at this wavelength. In Lemarchand (1992) we can find the scientific justification for the rest of *"magic frequencies."*<br><br>Noiseless region of the radio electromagnetic spectrum.<br><br>For laser nanosecond pulses there is no need to specify a frequency, we are looking for the number of arriving photons in a very small time slot. | The fact that the exploration of the cosmic environment need the deployment of radio astronomical studies in these specific wavelengths make the proposal interesting, because we assume that the galactic astronomers would eventually discover the "artificial signals" while looking for natural phenomena.<br><br>Perhaps an ETI with different "cognitive maps" would propose a different frequency with no relevant meaning for us. Specific equipment is built to explore the whole region of the spectrum (e.g *Project Phoenix* and *The Allen Telescope Array*).<br><br>With modest telescopes, we may detect a nanosecond laser pulse equivalent to the signals that our laser beams may generate. A specific transmitting strategy should be established for selecting the targets and deciding how much time they will be beaming into a specific star. To improve the detection odds a common synchronization strategy should be established. | The hydrogen line wavelength seems to be the most *"parsimonious"* proposal for a common communication channel (considering that the concept of "hydrogen" may be a *Cognitive Universal*) | |
| **Transmitting Output Power *(W)*** | 10. Depends on the ETI technological level of development, the transmitting strategy (e.g. omnidirectional, selected targets, microwave vs. laser beams, etc.), the characteristic of targets (e.g. Exoplanets, full survey, the use natural amplifiers as cosmic maser regions, etc.) | The assumptions on the transmitting power used by an ETI, defines the required sensitivity for the detection of the calling signals at a given *R* distance from us. Different strategies define different transmitting output power values. | We have the technological capability to detect a signal from the galactic center transmitted with the same equivalent isotropic radiated power available at the Arecibo radar. We can also detect laser signals from several hundred light years away. | From an aesthetical point of view the most "parsimonious" strategy will require a minimum output power in order to succeed. | From an ethical point of view the most "parsimonious" strategy will require a minimum output power in order to minimize the use of societal resources and keep the standards of environmental values. |
| **Polarization *(P)*** | 11. Elliptical<br><br>12. Linear | A narrowband signal is inherently polarized with the most general polarization being elliptical.<br><br>Linear polarization contains an additional unknown, namely the position angle of the field vec- | From a technological point of view, the artificial signals are highly polarized. It is relatively easy to distinguish | | |
| | | tors. Due to the interaction with the interstellar media, linear polarization seems unlikely. | the polarization characteristics with the present state-of-the-art technology. | | |
| | 13. Circular | Contains the least number of "unknowns" and is therefore the most likely. There are only two possible senses. | | | |
| **Modulation Type** ***(m)*** | 14. Amplitude | Since one cannot transmit a negative amplitude, and since a (0, 1) method is undetectable during the zero state and thus might be missed, we can probably rule out amplitude modulation. | | | |
| | 15. Phase | Binary phase modulation may be achieved by introducing a 180º phase shift into the transmitted signal at appropriate times. From an information carrying standard point, phase modulation is superior to the other methods. | However the signal a whole is not any easier to detect. Also phase modulation is least likely to be detected by serendipity, since astronomical observations rarely observe phase. So it would be important for the design and detection of high information content message but not for beacon signals. | | |
| | 16. Frequency | Is obtained by switching the transmitting frequency between two fixed frequencies at appropriate times. | | | |
| | 17. Polarization | May be achieved by changing between two orthogonal polarizations at appropriate times, such as two perpendicular linear polarizations or between left and right circular polarization | | | |
| | 18. Combinations | Perhaps, different systems of modulation would be used simultaneously in order to maximize the transmission of information and to deploy different levels of message complexities. | | | |
| **Signal bandwidth** ***(ΔB)*** | 19. Narrowband signals | For microwave transmissions, the interstellar scintillations set a limit to the possible artificial narrowband signals (≈1 Hz). An ultra-narrowband signal would be the most distinctive characteristic of an artificial signal. | From a technological point of view the transmission of an ultra narrowband signal for the generation of a "beacon" has the advantage that the communication range would be larger than using a broadband signal with the same output power. | | |
| | 20. Broadband signals | Nanosecond Laser beams has a broadband characteristic, and the artificial characteristic would be established by the nanosecond variation of the flash-light. | Broadband signals are useful to send huge amounts of information. From a microwave detection point of view, these signals would be use only after the contact is already established. From the laser detection point of view the present technology allow us to send and received enormous amounts of information using this communication channel. | | |
| **Signaling rate** ***($\tau_l$)*** | 21. Beacons (CW signals) | Low information rate. The only information would be the location of the artificial source and the transmitting power. | The easier message to detect, transmit and decoded. | | |
| | 22. Small duty cycle (low information content) | The interstellar scintillation place a limit to microwave information rates into a few thousand bits per second. | These transmissions are easy to generate to detect using the state-of-the-art technologies. It may be difficult to decode and interpret the signals. | | |
| | 23. Nanosecond and subnanosecond high information content message | Using laser beams it is possible to exchange large amounts of information across interstellar distances. | | | |
| | 24. Beacons | No decoding is needed. No specific message is included | | | Possible solution of the *LexGal* for those transmissions in the direction of unknown recipients. |
| **Code** ***(C )*** | 25. Binary code | The most effective method for maximum range communication is the two state (binary) one. This way of coding is the one that may follow the anti-cryptography principle. | | This method may be useful to be employed to send some “aesthetical cognitive universals” (e.g. *golden cuboid*) | This method may be compatible with *LexGal* for the case of sending, aesthetical, ethical or low scientific message contents. |
| | 26. Complex codes | It is possible to imagine very sophisticated coding systems, but in these cases it would be practically impossible to be decoded by emergent societies. | | | This method may be used among advanced technological societies or just to send the first chapters of the *Encyclopedia Galactica.* |
| | 27. Logical | Organize the information in a sequence of increasing complexity, using logical principles, like in LINCOS. Another option is by creating numerically coded symbols and explicitly describing their context within an expression. | Some attempts where already made to develop a specific semantics by Freudenthal, Devito and others. | | This method is useful for deploying the *Encyclopedia Galactica* or a high information content message. According to *LexGal*, only expected from nearby ($R<$ 70 ly) civilizations, which already have calibrated our level of ethical development. |
| **Semantics** ***(S)*** | 28. Analogical | Pictographic communication systems: The representation of images or holograms that represent objects or actions creating analogies to similar objects or actions that our cognitive map would recognize or find it familiar. | By 1820, K. Gauss proposed to represent the image of the geometrical demonstration of the Theorem of Pythagoras to hypothetical inhabitants of the Moon. Similar ideas were proposed over the years to communicate with the Martians or with other stars. Drake and others optimized the proceeding and in 1971 sent a message in this form to the globular cluster M31. In the late ’90 Dutil and Dumas sent similar semantics messages to a group of nearby stars. | This method would be useful to communicate different forms of visual arts, as well as music. | This method is appropriated to send aesthetical messages, according to *LexGal* ($R>>$ 70 ly) |
| | 29.Heuristic-Algorithm | Binary computing: the binary code may represent logical instructions for computer processing | This proposal requires the transmission of a complex software that may be useful to communicate concepts in a great variety of forms and eventually also interact in some way with the message recipients. | | This procedure would require large amounts of information contents messages, proportional to the complexity of the “software”. Eventually, for the case of the interacting versions, the program may be designed in such a way in order to avoid complex information to low level of ethical development societies. So some “smart” algorithms would work following the *LexGal*. |
| **Kind of Information Content *(KIC)*** | 30. Beacon | To indicate the location of an advanced technological civilization in the galaxy and to communicate "that we are not alone". | From a technological point of view this is the easiest message to be detected (e.g. CW monochromatic signal). | | Using the *LexGal* and the *PUF*, the transmitting civilization that ignores the level of ethical development of the recipients, would chose this strategy until establishing which knowledge would not put in danger the natural evolution of the recipient civilization. |
| | 31. Scientific & Technological | Information about nature and evolution of the universe. Description of scientific conceptualizations in order manage all our environmental variables, e.g. to travel across wormholes, to discover new energy sources, to learn more about complex systems (life, societies, etc.), recipes to avoid self-destruction, etc. | Description on how to build a new and more efficient communication system, travel machine or some sort of interacting machine to exchange information with it. | | Scientific and technological information would be share only in the case of similar evolutionary stage between sender and recipient. Applying the *LexGal*, the sender will start with those chapters of the *Encyclopedia Galactica* that will not put in danger of recipient self-annihilation due to the misused of the message content. This strategy may be the one that a hypothetical nearby advanced civilizations (*R*<< 70 ly) will apply to us, giving us only those things that we can manage. |
| | 32. Artistic | Incommensurability of artistic cognitive maps? | | The diversity of culture, what makes a society different from another one is probably the most important information to be shared with others. Visual arts, music and an unknown complete different conception of art may be the most probable high-information content of an interstellar message. The learning process of new art forms will induce us to generate "new" cognitive maps of nature. The history of science and arts on Earth shows how this feedback between these two cultural activities modified our human cognitive maps. | Using *LexGal* an artistic content would not put in danger of self-annihilation to the recipient society. The learning process to perceive the new forms of arts would help the recipients to develop new and different cognitive maps that would help them to be prepared for receiving the next "chapters" of the *Encyclopedia Galactica*. |
| | 33. Ethical | Ethical information content would be very important for the development of the so-called "social sciences". A galactic cross study about the evolution of ethical systems in different biological systems would be of an enormous importance to learn how to improve our human ethical behavior in time. | Ethical information content will be of high importance in order to develop "disembodied technologies" to solve several of the most important problems of our present human society. | For the ancient Greeks, there was a strong connection between ethics and aesthetics. For them any ethical solution for a human problem was beautiful. Present studies about the evolution of altruism in different species shows the appearance of interesting patterns and symmetries between the members of the "societies" under study.<br>A hypothetical ethical message content may also present unexpected aesthetical patterns that would also help us to develop new cognitive maps. | Ethical information content may also help to improve our level of ethical development, so this KIC may be an excellent instrument of the *LexGal* argument and the *PUF*. |
| | 34. Metaphysical-Religious | Is God a cognitive universal? As it was the case of early human cognitive maps, metaphysical and religious conceptions of nature were considered part of the | | | |
| | | epistemological explanation about the origin and evolution of the universe. We cannot rule out completely the possibility that an ETI will send their own metaphysical interpretation of nature. | | | |